\documentclass[submitting]{nst}
\usepackage{comment}
\usepackage{subfigure,dcolumn}
\usepackage[T2A,T1]{fontenc}
\usepackage[russian,english]{babel}
\usepackage{booktabs}
\usepackage{listings}
\begin{document}

\title{Performance Evaluation of a High‑Granularity LYSO–SiPM Based Position‑Sensitive Detector for a One-Shot $\gamma$-Scanning System with Sub‑Millimeter Spatial Resolution} 

\author{Katyayni Tiwari}
\email[Corresponding author, ]{pps@iitrpr.ac.in}
\affiliation{Department of Physics, Indian Institute of Technology Ropar, Rupnagar 140001, Punjab, India}

\author{Arzoo Sharma}
\email[Corresponding author, ]{arzoo@gauhati.ac.in}
\affiliation{Department of Physics, Gauhati University, Guwahati 781014, Assam, India}

\author{J. Gerl}
\affiliation{GSI Helmholtzzentrum für Schwerionenforschung, Darmstadt 64291, Germany}

\author{I. Kojouharov}
\affiliation{GSI Helmholtzzentrum für Schwerionenforschung, Darmstadt 64291, Germany}

\author{P. Herrmann}
\affiliation{GSI Helmholtzzentrum für Schwerionenforschung, Darmstadt 64291, Germany}

\author{H. Schaffner}
\affiliation{GSI Helmholtzzentrum für Schwerionenforschung, Darmstadt 64291, Germany}

\author{G. Aggez}
\affiliation{Istanbul University, Department of Physics, Istanbul 34116, Turkey}

\author{Pushpendra P. Singh}
\email[Corresponding author, ]{pps@iitrpr.ac.in}
\affiliation{Department of Physics, Indian Institute of Technology Ropar, Rupnagar 140001, Punjab, India}

\begin{abstract}
A compact position-sensitive $\gamma$-detector based on a thin monolithic LYSO crystal and a 96-channel SiPM array is developed for the spatial characterization and calibration of segmented $\gamma$-ray detector systems used in nuclear-physics experiments. The detector is designed to provide localized irradiation and rapid two-dimensional response mapping. A 7~cm diameter and 3~mm thick LYSO crystal is optically coupled to the SiPM array, and the position of the incident $\gamma$-ray interaction is reconstructed from the relative scintillation-light signals collected by neighboring SiPM channels. An asymmetry-based charge-sharing method is employed to determine the interaction position from the spatial distribution of the detected scintillation light. Detailed GEANT4 simulations, including optical photon transport, were performed to investigate the detector response and to estimate its intrinsic spatial resolution. The simulations predict a spatial resolution of approximately 0.5~mm for 60~keV and 511~keV $\gamma$-rays under idealized conditions. A prototype detector was developed and experimentally characterized using coincidence measurements with the $\gamma$-scanning facility at GSI, Germany. The experimental measurements demonstrate a spatial resolution better than 1 mm in the central detector region, while the position-dependent response and degradation near the detector boundaries are investigated. The experimental results are compared with GEANT4 predictions to identify the contributions of optical photon transport, charge sharing, and detector geometry to the measured spatial resolution. The developed detector provides a compact, high-gain advanced solution for 3D characterization of highly segmented $\gamma$-ray detector arrays in nuclear physics experiments.

\end{abstract}


\maketitle
\nolinenumbers

\section{Introduction}

Gamma-ray spectroscopy has advanced considerably with the development of highly segmented, position-sensitive germanium detectors, such as the Advanced GAmma-ray Tracking Array (AGATA) \cite{eberth2008ge, eberth2023agata, korten2020physics}. These detectors employ pulse-shape analysis (PSA) to determine the energy and three-dimensional interaction positions of individual $\gamma$-ray events, enabling accurate $\gamma$-ray tracking and reconstruction \cite{venturelli2004adaptive, crespi2007, krammer2021tracking}. Reliable tracking requires a detailed characterization of the detector response over its active volume. In particular, a comprehensive database of experimentally measured and simulated pulse shapes at different interaction positions is essential for developing and validating PSA algorithms. Dedicated scanning systems, such as the Liverpool scanner, have therefore been developed to characterize segmented detector crystals in three dimensions using a collimated radioactive source \cite{dimmock2009validation, domingo2011novel, habermann2017}. However, scanning a large number of positions over a detector volume is highly time-consuming; for example, characterization of approximately 1200 points in an AGATA crystal required about two months \cite{goel2013, GOEL2011}. This highlights the need for compact and efficient position-sensitive detectors capable of providing rapid three-dimensional characterization of segmented $\gamma$-ray detector systems. A position-sensitive detectors are also essential for detector characterization, calibration of segmented detector arrays, and the development of next-generation $\gamma$-ray tracking arrays \cite{du2020detector}
To address this requirement, a compact position-sensitive $\gamma$-ray detector is being developed jointly by IIT Ropar and GSI, Germany, using a monolithic cerium-doped lutetium–yttrium oxyorthosilicate (LYSO:Ce) scintillator coupled to a 96 silicon photomultiplier (SiPM) matrix. The choice of scintillator material plays a central role in achieving the desired detector performance. Some inorganic scintillators, such as NaI(Tl), BGO, and LYSO, exhibit different trade-offs among stopping power, light yield, decay time, and mechanical robustness. Among these, LYSO has emerged as a preferred material for modern PET systems due to its high density, fast decay time, and high light yield, enabling excellent timing and energy resolution \cite{singh2024review, wei1501intrinsic, melcher1992lyso}. In addition, LYSO is non-hygroscopic, making it well-suited for compact detectors \cite{Enriquez2020}. The choice of photodetector for scintillation light readout is equally important. Silicon photomultipliers (SiPMs) are increasingly replacing traditional position-sensitive photomultiplier tubes (PSPMTs) due to their compact size, high gain, low operating voltage, cost-effectiveness, and insensitivity to magnetic fields \cite{otte2005test, roncali2019SiPM}. SiPM-based readout systems have demonstrated significant improvements in both spatial resolution and timing performance, making them a suitable choice in modern PET and $\gamma$-imaging detectors \cite{roncali2019SiPM, seifert2012timing}. With a 7 cm crystal diameter and a SiPM configuration, the detector is designed to provide a large field of view while retaining high spatial sensitivity. The resulting system offers a compact approach for rapid spatial characterization and calibration of segmented $\gamma$-ray detector arrays, with potential applications in $\gamma$-ray imaging, spectroscopy, and detector development.
In this work, we present the detector design, GEANT4 simulation, and experimental characterization of the LYSO--SiPM detector. Its spatial response is investigated under different irradiation conditions, followed by position reconstruction and quantitative evaluation of the achieved spatial resolution. Preliminary results were reported in \cite{tiwari2024new}; the present work provides a detailed study of the detector design, simulation, spatial sensitivity, and reconstruction performance.
\par
\vspace{-5.5mm}
\section{Detector Description}
\begin{figure}[!ht]]
	\centering	\includegraphics[width=8.2cm]{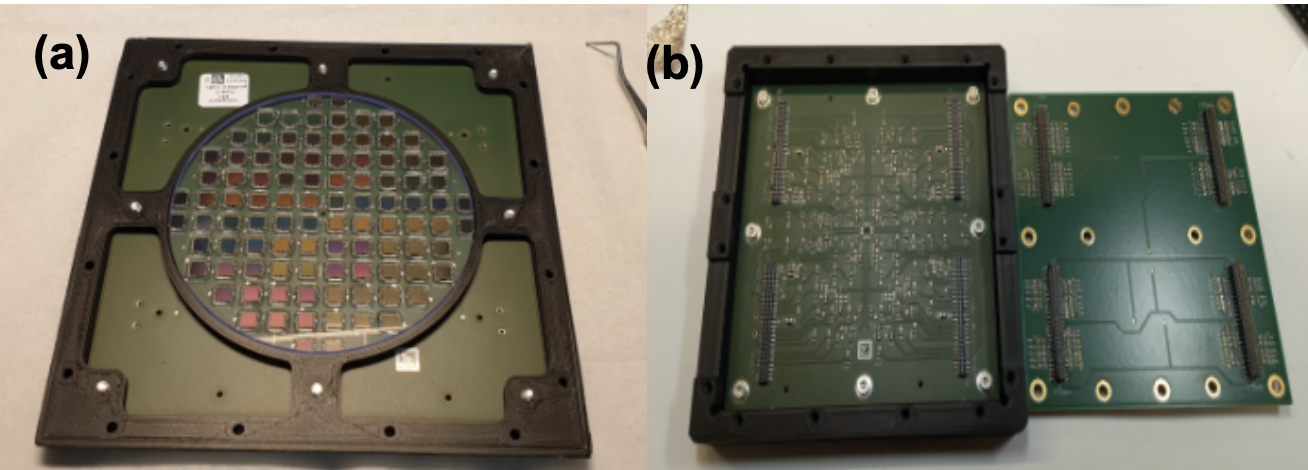}
	\caption{Key development stages for compact PSD assembly. (a) Cylindrical LYSO scintillator with reflective glass coupled to SiPM board (green PCB). An outer black body is used to secure the scintillator crystal and fully enclose the assembly. (b) Front view showing the SiPM matrix readout.}
	\label{fig1} 
    \end{figure}

\vspace{-3mm}
The LYSO crystal has a diameter of 7~cm and a thickness of 3~mm, providing a compact geometry for the detection of $\gamma$-rays while reducing parallax effects. The array has been constructed on a PCB with a back-end containing an RC circuit for pulse readout, i.e., reading the electrical pulse produced after an electron-hole pair is generated by scintillation light falling on the SiPM. Figure \ref{fig1}(a) shows the LYSO scintillator coupled to the SiPM matrix, and Figure \ref{fig1}(b) shows the SiPM readout circuit. The scintillator has been covered with thin Aluminized Mylar foil to increase reflectivity. Lastly, the entire assembly has been enclosed in a tightly packed black box. To further block outside light, black tape (made of polyester that withstands temperatures from -40 to 121°C) has been wrapped around the entire detector geometry.
The chosen layout represents a balance between high spatial granularity (for better resolution), reduced dead space (for efficiency), and manageable readout channels (for cost and complexity). Optical coupling between LYSO and the SiPM array ensures efficient light collection. The detector design aims to achieve high photon detection efficiency and accurate position reconstruction.

\section{GEANT4 Modeling to Calculate Intrinsic Position Resolution}

The simulation was performed to evaluate the intrinsic spatial resolution of the detector. Electronic noise and signal processing effects were not included at this stage. The focus was on understanding the fundamental performance limits of the detector system \cite{moszynski2002properties}. Detector response was simulated using GEANT4, a versatile simulation toolkit built using object-oriented programming \cite{wang2024optical, allison2006geant4, dev2024assessing}. It provides a comprehensive framework to simulate particle transport, electromagnetic processes, and optical photon propagation with high accuracy \cite{dietz2016peculiarities, khodaei2023review}. 

\subsection{Optical Properties and Boundary Conditions}

The optical response of the detector was modeled in GEANT4 by assigning wavelength-dependent optical properties to the LYSO crystal, optical coupling medium, and SiPM interface. The LYSO optical properties included its refractive index, bulk optical absorption/transmittance, and surface reflectance. The optical glue was modeled as an optical coupling layer with its corresponding refractive index and reflectance. At the SiPM interface, an optical boundary was defined between the coupling medium and the SiPM entrance surface, allowing photons to be transmitted, reflected, or detected according to the specified interface properties. These parameters were included to account for Fresnel reflection and photon losses at material interfaces and to reproduce the measured light-collection response, as mentioned in Table~\ref{tab:sipm_optical_surface}.

\begin{figure}[h!]
	\centering
	\includegraphics[width=8cm]{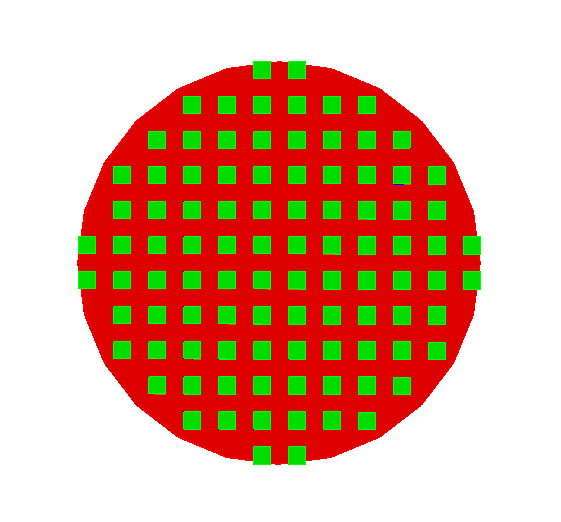}
	\caption{GEANT4 simulated back view of the detector design comprising a LYSO (in red) crystal coupled to a 96-SiPM (in green) array.}
	\label{fig3}
\end{figure}

The optical photon propagation through the crystal introduces statistical errors that follow the Landau-Gaussian distribution. G4EmStandardPhysics\_option4 and G4OpticalPhysics were used for the simulation. An air layer was configured between the crystal and the Enhanced Specular Reflector (ESR) layer. At the same time, a dielectric-dielectric surface type was specified for the interfaces between air and silicone and between air and LYSO. Similarly, the dielectric-metal surface type was used at the boundaries between ESR and air and between ESR and silicone. Additionally, for the ESR boundary, the photoelectric efficiency was set to 0 and the reflectivity to 0.985, while for the LYSO boundary, the surface polishing was adjustable within the range of (0, 1). The refractive index of optical glue has also been considered between the LYSO and SiPMs \cite{wang2024optical}.
Relevant physical processes, such as $\gamma$-ray interactions, scintillation light generation, and optical photon transport, were included. Optical properties of materials, such as refractive index, absorption length, and scintillation yield, were defined to ensure realistic light propagation within the detector and implemented according to the manufacturer's Table of Optical Properties of Silicon \cite{jo2023erratum}. The crystal geometry used in the simulation is identical to that of the real detector described in the previous section. The GEANT4-simulated detector is illustrated in Figure~\ref{fig3}, which shows LYSO coupled to SiPMs.

\begin{table*}[htbp]
\centering
\caption{Optical surface properties used to model the SiPM boundary in the GEANT4 simulation.}
\label{tab:sipm_optical_surface}

\begin{tabular}{lll}
\toprule
\textbf{Parameter} & \textbf{Value in simulation} & \textbf{Physical meaning} \\
\midrule

Optical surface
& \texttt{SiPM\_opsurf}
& Optical boundary assigned to the SiPM \\

Surface model
& GLISUR
& Original Geant3-based optical surface model implemented in Geant4 \\

Surface finish
& Polished
& Smooth/specular surface; no rough or ground treatment \\

Surface type
& \texttt{dielectric\_metal}
& Optical photons incident on SiPM as interacting with a metal-like surface \\

Surface assignment
& \texttt{G4LogicalSkinSurface}
& Surface is applied to all surfaces of the SiPM logical volume \\

Surface MPT
& \texttt{myMPT5}
& Contains the optical response as a function of photon energy \\

Detection efficiency
& \texttt{EFFICIENCY = 1}
& Every photon absorbed at the surface is eligible to be detected \\

Explicit reflectivity
& \texttt{REFLECTIVITY = 0}
& No user-defined reflected fraction is specified through this property \\

Complex refractive index
& \texttt{REALRINDEX}, \texttt{IMAGINARYRINDEX}
& Used to characterize the optical response of the SiPM surface \\

\bottomrule
\end{tabular}
\end{table*}

\subsection{Simulation Methodology and analysis}
Mono-energetic pencil beams of 60~keV and 511~keV $\gamma$-rays were used to irradiate the detector at predefined positions perpendicular to its front face. This allows precise evaluation of position reconstruction performance without spatial spread from the source, providing ideal beam conditions for the detector. A total of 5.$10^5$ events were simulated for each irradiated position to ensure sufficient statistical accuracy. A large dataset was used to reduce statistical fluctuations and improve the reliability of the extracted spatial resolution. For each event, the number of optical photons detected by individual SiPMs was recorded. This event-wise photon distribution forms the basis for position reconstruction and further analysis. The interaction position was reconstructed using the barycenter method, which computes the positions as a weighted average of the SiPM coordinates. The interaction position was reconstructed using the barycenter method defined as: 

\begin{equation}
\label{eq:barycentre1}
\bar{x} = \frac{\sum_{i} x_i \, w_i}{\sum_{i} w_i}
\end{equation}

\begin{equation}
\label{eq:barycentre2}
\bar{y} = \frac{\sum_{i} y_i \, w_i}{\sum_{i} w_i}
\end{equation}

Where $\bar{x}, \bar{y}$ is the reconstructed interaction position along the x- and y-axes, respectively. $x_i, y_i$ is the physical coordinate of the $i$-th SiPM in the detector array. $w_i$ is the number of detected photons recorded by the $i$-th SiPM. $\sum_{i} w_i$ is the total number of photons detected across all SiPMs for a given event. 

\begin{figure}[h!]
	\centering
	\includegraphics[width=8.7cm]{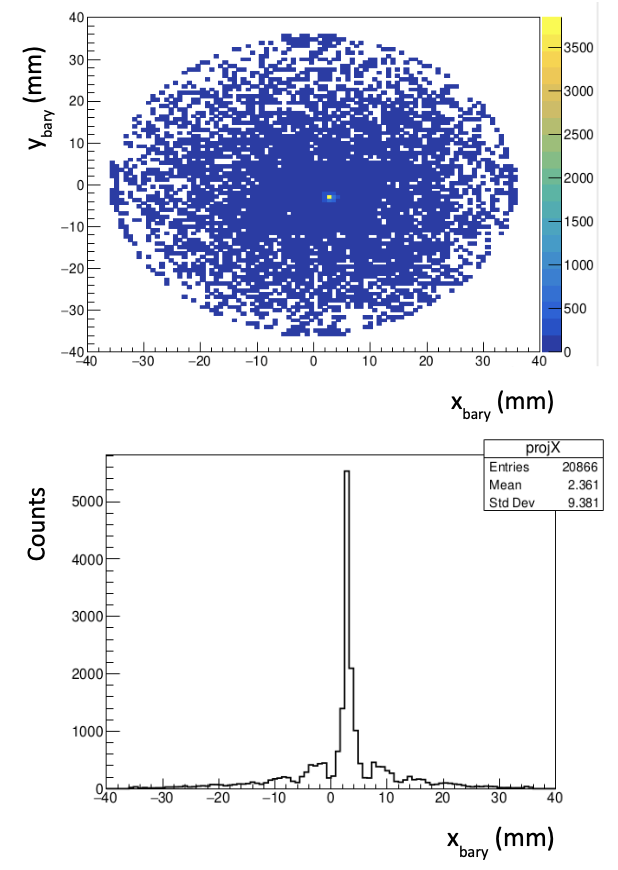} 
	\caption{GEANT4-simulated 2D distribution of photon hits for a 511~keV pencil $\gamma$-beam incident at the center of the SiPM array using the barycenter approach using Equation~(\ref{eq:barycentre1} \& \ref{eq:barycentre2}). The heatmap highlights a pronounced peak in photon counts at the detector center, with a smooth radial decrease toward the periphery. The corresponding x-projection shown below displays narrow, symmetric peaks centered at zero, consistent with the expected detector response.}
	\label{fig4}
\end{figure}

This method assigns greater importance to SiPMs that detect more photons, thereby biasing the reconstructed position toward regions of maximum light intensity \cite{moehrs2006detector}. The reconstructed positions were used to generate a 2D heatmap of detector response. This provides a visual representation of the spatial distribution of reconstructed interaction points. The reconstructed distributions exhibit a sharp, localized peak that resembles a Dirac delta in the projection plots shown in Figure~\ref{fig4}. This indicates the good localization capability of the detector.

\vspace{-9mm}
To evaluate spatial resolution, systematic scans were performed along the +$x$ and +$y$ axes. The beam position was varied in 0.4~mm steps. An asymmetry parameter (asym\_param) was calculated based on differences in photon counts across selected SiPM groups using Equation~(\ref{eq:asym_param}). This parameter provides a sensitive measure of position variation across the detector. A linear fit was applied to extract the slope, which gives information related to position sensitivity.
The analysis was carried out for both the central and edge regions of the detector. The detector achieves an intrinsic spatial resolution of approximately 0.5 mm, demonstrating sub-millimeter performance under ideal conditions, as mentioned in Table \ref{tab:position_resolution_combined}. It should be noted that the simulation does not include electronic effects such as noise, gain variation, or timing jitter. Only the quantum efficiency of the SiPMs was considered in the light-photon statistics. Therefore, the reported resolution represents the best achievable intrinsic resolution of the detector. These results provide a baseline for comparison with experimental measurements discussed in the next section.
\section{Experimental Section}
\subsection{Channel Calibration and Data Acquisition}
\begin{figure}[htbp]
    \centering
    \includegraphics[width=0.45\textwidth]{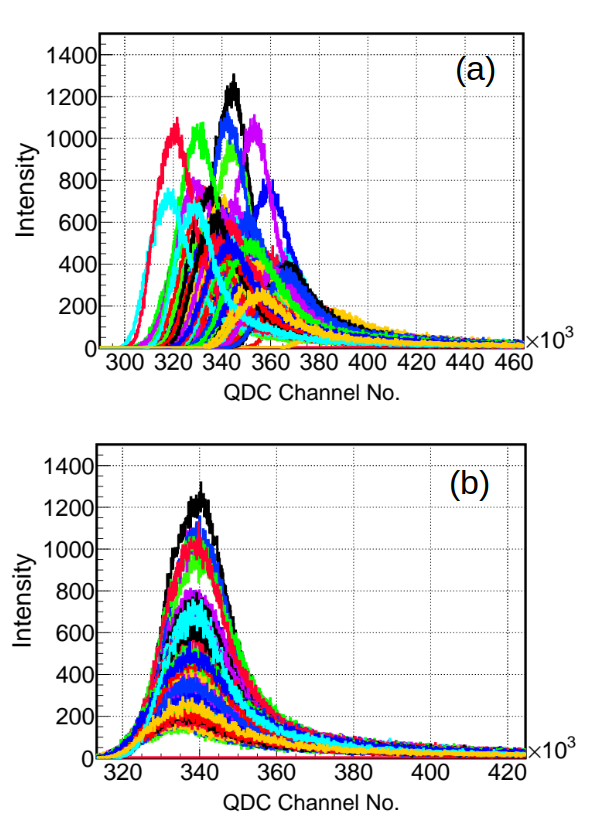} 
    \caption{(a) TAMEX-TOT spectra for all 96 channels (before gain matching), and (b) TAMEX-TOT spectra after the gain matching procedure for the coincidence setup between the GSI scanner and the newly developed imager.}
    \label{qdc}
    \vspace{-3mm}
\end{figure}

In the first measurement setup, the background data from an overnight run lasting approximately 12 hours was used to gain-match the channels. The channels before and after gain matching are shown in Figures~\ref{qdc}(a)-(b). The matching is performed by subtracting the pedestal value (noise in the QDC channel).

The experimental setup was developed to validate the detector performance under controlled conditions as follows:

\begin{itemize}
    \item A VME-based 32-channel QDC was used for the GSI scanner. The GSI scanner serves as the reference/characterization detector used to evaluate the performance of the newly developed PSD \cite{crawford2017nuclear, goel2013}. The detector was connected to a TAMEX FPGA-based digitizer. For the full 96-channel PSD readout, six TAMEX cards, each with 32 input channels, were used, along with a front-end twinpeak readout board with 16 channels. TAMEX, a Time-over-Threshold (TOT) digitization system, was chosen to provide a simplified and cost-effective acquisition system with excellent timing precision of about 15 ps \cite{benetti2011simulation}.
    
    \item Before data acquisition, an online fine-time calibration was conducted using the internal pulser of the TAMEX system via the Multi Branch System (MBS) terminal command line \cite{MBS}. This procedure was performed to consolidate all channels into a common binning scheme. Additionally, threshold settings were applied to each channel, and the trigger was set for all channels. Multiple coincidence measurements were made by adjusting the threshold level in TAMEX. The results of one of the optimized threshold settings are presented in this work. The PSD bias voltage was set to -30 V. The coincidence data rate was $\approx 100$ Hz. 
\end{itemize}

\begin{figure}[h!]
\centering
\includegraphics[width=8.8cm]{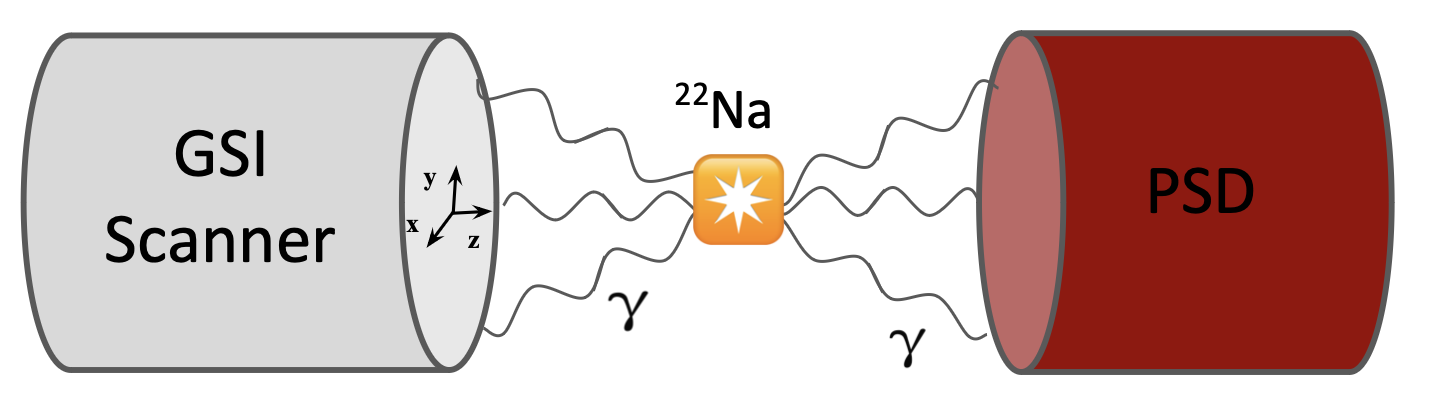}
\caption{Schematic of the coincidence detector setup featuring the GSI scanner on the left and a new position-sensitive detector (PSD) on the right, with a \texorpdfstring{$^{22}$Na}{22-Na} source. The \texorpdfstring{$^{22}$Na}{22-Na} source is kept at a distance of 8.2~cm from the PSD. In this configuration, the GSI scanner serves as the reference detector, and the PSD serves as the triggered detector.}
\label{fig5}    
\end{figure}

Further, 511~keV $\gamma$-rays from a $^{22}$Na source were used for measurements. The detector assembly includes a LYSO crystal, a 96-SiPM array, and a readout electronics system. The distance between the source and the detector was $\approx 8.2$ cm. The coincidence detector setup is represented in Figure~\ref{fig5}. This distance was optimized to ensure full solid-angle coverage around the detector. The optimization was carried out by monitoring 2D images in real time using the GO4 software \cite{GSI-Go4}. 

\section{Experimental Analysis}
To evaluate the performance of the developed position-sensitive detector, it was operated in a coincidence configuration by integrating it into the GSI $\gamma$-scanning system. The same $^{22}$Na source was used, enabling the detection of back-to-back 511 keV annihilation photons. This arrangement enabled evaluation of the detector’s spatial response using imaging techniques, providing insight into its full performance. The coincidence setup significantly improved background suppression, thereby enhancing the accuracy and reliability of the measured detector parameters.

\subsection{Coincidence test between the GSI scanner and the PSD}

\begin{figure*}[!htb]
\centering
\includegraphics[width=17cm]{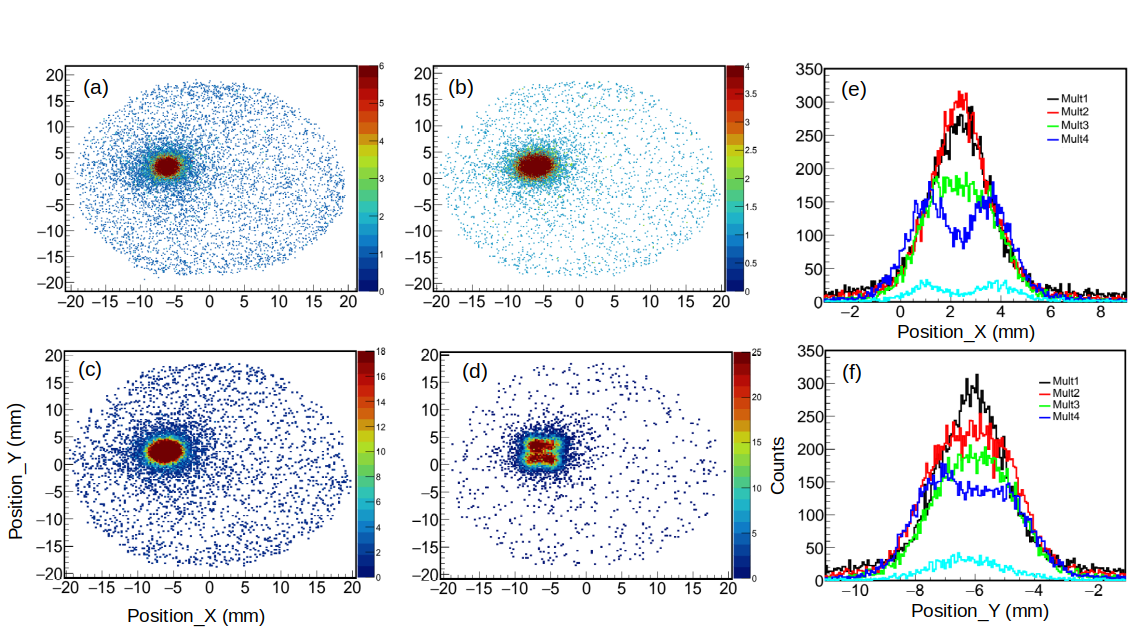}
\caption{A 2D representation of various interaction possibilities for different SiPM hit multiplicities for reference SiPM, S26. Panels (a)-(d) illustrate multiplicities 1 - 4, respectively. The initially circular shape of the hit SiPM in multiplicity one gradually deforms into a square-like topology as the multiplicity increases. Panels (e)–(f) show the projected distributions of interaction probabilities for event multiplicities ranging from 1 to 4, along the x-axis and y-axis, respectively. The initial Gaussian distribution for multiplicity one in both axes gradually transforms into two flattened Gaussian peaks, reaching this shape by multiplicity 4.}\label{fig7}    
\end{figure*} 

A 2D heatmap of reconstructed positions was generated to show the detector response. Projection plots along the x and y axes were used to evaluate spatial resolution, as shown in Figure~\ref{fig7}. It shows the reconstructed interaction patterns for different SiPMs multiplicities of one of the reference SiPMs, S26, with multiplicities 1-4 displayed in panels (a-d), respectively. The initially circular distribution becomes square-like at higher multiplicities, indicating increased charge sharing between neighboring SiPM channels. These are the 2D images obtained using the GSI $\gamma$-scanner. Furthermore, projection images along the $x$ and $y$ axes, shown in panels (e) and (f), illustrate the distribution of interactions for multiplicities ranging from 1 to 4. The initial Gaussian distribution seen for multiplicity 1 in both axes gradually evolves into two flattened Gaussian peaks for multiplicity 4. The detector shows a localized response consistent with simulation trends. Position resolution was calculated using the standard deviation ($\sigma$) from projection distributions.

\begin{figure*}[!htb]
    \centering
    \includegraphics[width=0.75\textwidth]{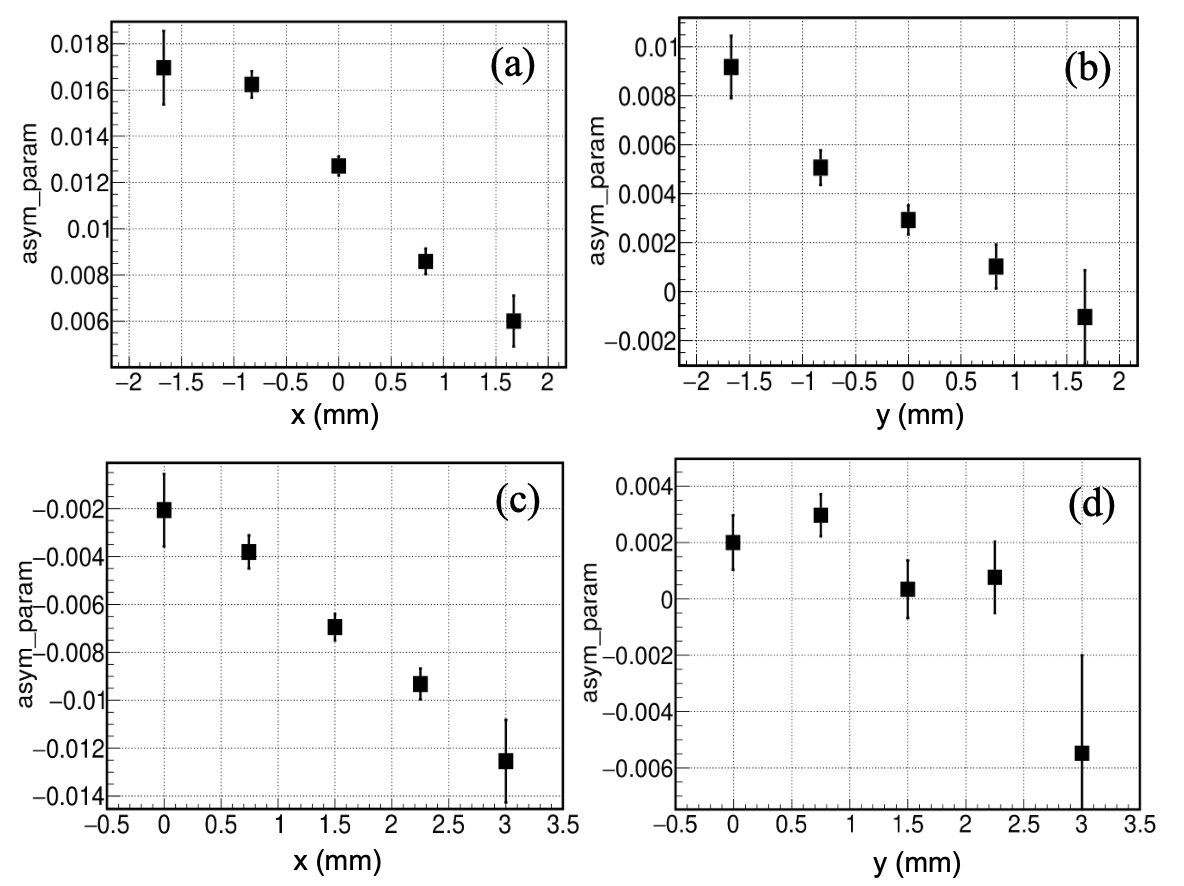}
    \caption{Asymmetry parameter (asym\_param) at 511 keV derived from experimental measurements of $\gamma$-interactions shown using central SiPM, S26 as a reference channel along the (a) $x$-axis and (b) $y$-axis, and a edge SiPM S85 along the (c) $x$-axis, and (d) $y$-axis respectively. The slope of these curves is used to obtain the position resolution listed in Table~\ref{tab:position_resolution_combined}.}
    \label{fig8}
\end{figure*}

To quantify this position dependence and investigate charge sharing between neighboring channels in the PSD, an asymmetry parameter (asym\_param) is defined. This parameter is constructed from the ratio of charges measured in the left- and right-neighboring SiPMs, providing a measure sensitive to shifts in the interaction position while reducing the influence of position \cite{arzoo, Cooper2008}.

The asymmetry parameter is defined as:
\begin{equation}
\label{eq:asym_param}
asym\_param = \frac{A_R - A_L}{A_R + A_L},
\end{equation}
\\
where $A_R$ and $A_L$ are maximum amplitude values in the neighboring right and left segments of the hit segment, respectively, the asym\_param for 511 keV was derived from experimental measurements of $\gamma$-interactions. It was evaluated using the central SiPM S26 as a reference SiPM along the (a) $x$-axis and (b) $y$-axis, and the edge SiPM S85 along the (c) $x$-axis and (d) $y$-axis, as shown in Figures~\ref{fig8} (a-d), respectively. 

The asym\_param has been calculated for various x-axis cuts. For instance, for the reference SiPM S26, the neighboring channels along the $x$-axis are S25 and S27. To calculate the asym\_param for S26, the amplitude values of these neighboring SiPMs are applied in Equation~(\ref{eq:asym_param}). Along the $y$-axis, the asym\_param was calculated in the same manner as along the $x$-axis. A linear fit to obtain asymmetry plots was used to determine the position resolution using a weighted mean calculation. For the $x$-axis, the resolution is of the order of $\approx$ 0.9 $\pm$ 0.3 mm, and for the $y$-axis, it is $\approx$ 0.8 $\pm$ 0.2 mm. Further, position resolution was calculated for some additional reference SiPMs, S52, S85, and S62, along the $x$ and $y$-axes, as listed in Table~\ref{tab:position_resolution_combined}.

\begin{table*}[t]
\caption{Position resolution along the $x$ and $y$ axes for 60~keV and 511~keV $\gamma$-energies. Results are shown for both GEANT4 simulations and experimental measurements at different SiPM positions.}
\label{tab:position_resolution_combined}
\centering

\renewcommand{\arraystretch}{1.8}
\setlength{\tabcolsep}{10pt}

\small
\begin{tabular}{ccc}
\hline
Energy (Type) & Hit SiPM (Position) & Resolution (mm) \\
\hline

60 keV (Sim) & S26 (center) & $x:~0.38 \pm 0.30,\; y:~0.46 \pm 0.35$ \\
             & S2 (edge)    & $x:~0.44 \pm 0.34,\; y:~0.43 \pm 0.33$ \\

\hline

511 keV (Sim) & S26 (center) & $x:~0.37 \pm 0.21,\; y:~0.43 \pm 0.23$ \\
              & S2 (edge)    & $x:~0.42 \pm 0.28,\; y:~0.39 \pm 0.26$ \\

\hline

511 keV (Exp) & S52 (edge)   & $x:~0.86 \pm 0.29$ \\
              & S26 (center) & $x:~0.90 \pm 0.30,\; y:~0.80 \pm 0.21$ \\
              & S64 (edge)   & $y:~0.79 \pm 0.23$ \\
              & S85 (edge)   & $x:~0.76 \pm 0.18$ \\
              & S62 (edge)   & $x:~0.88 \pm 0.15$ \\

\hline
\end{tabular}
\end{table*}

\section{Experimental results and comparison with simulations}
The simulation and experimental results, as listed in Table~\ref{tab:position_resolution_combined} show consistent spatial-response trends, with the experimental resolution remaining below 1 mm. The difference in absolute resolution is attributable primarily to non-ideal effects present in the experimental readout chain but not included in the GEANT4 model.
It shows good agreement between simulation and experimental position-resolution values, validating the detector design. Slight deviations arise from real-world factors such as electronic noise, SiPM gain variations, and optical photon losses. \\
GEANT4 result: intrinsic detector/optical response under idealized conditions \\
Experimental result: end-to-end system resolution.

\section{Conclusion and future work}
The R\&D has been carried out to develop a position-sensitive $\gamma$-detector based on LYSO and 96 SiPMs, which has been reported in the present work. The optimized combination of scintillator characteristics, SiPM layout, and detector dimensions enables sub-millimeter position resolution via a barycentre reconstruction method, highlighting the effectiveness of this detector configuration. GEANT4 simulations demonstrate an intrinsic position resolution of $<0.5$~mm. Experimental results confirm the detector’s capability with comparable performance trends at the $<1$~mm spatial resolution.  \\

In the present work, the data analysis was performed using an analytical reconstruction approach, which provides a systematic framework for interpreting the detector response. However, further improvements in spatial resolution and reconstruction accuracy can be achieved using machine–learning–based methods. In future work, GEANT4-generated simulation datasets will be integrated into a machine learning framework to train models to predict the position of the $\gamma$ interaction within the detector. Such data-driven approaches are expected to enhance spatial accuracy, improve reconstruction robustness, and provide a more efficient analysis pipeline for position-sensitive detector systems. 

\vspace{2mm}
\section{Acknowledgement}
\vspace{-3mm}
The authors acknowledge GSI Helmholtzzentrum f\" {u}r Schwerionenforschung, Germany, for providing access to their experimental facility, all members of the GSI/DESPEC (DEcay SPECtroscopy) group for their help and valuable feedback, and the Indian Institute of Technology Ropar for an ISIRD Grant for the development of the scanner. The authors also acknowledge Dr. Helena May Albers, Dr. Tobias Habermann, Dr. Biswarup Das, and Dr. Magdalena Górska for their help and experimental support. One of the authors, K.T., thanks the Department of Science \& Technology for the INSPIRE Fellowship, and A.S. acknowledges GSI GET\textunderscore INvolved/FAIR for financial support.

\bibliographystyle{unsrt}
\bibliography{mybibliography}

\end{document}